\documentclass[pra,twocolumn,showpacs,amsfonts,amsmath,floatfix]{revtex4}
\usepackage{epsfig,amsmath,graphicx}

\usepackage{graphicx}% Include figure files
\usepackage{dcolumn}% Align table columns on decimal point
\usepackage{bm}% bold math
\usepackage{hyperref}% add hypertext capabilities
\begin{document}

%\preprint{APS/123-QED}

\title{Influence of spin dependent carrier dynamics on the properties of a dual-frequency vertical-external-cavity surface-emitting laser}% Force line breaks with \\
%\thanks{A footnote to the article title}%

\author{Syamsundar De$^{1}$}
\author{Vladyslav Potapchuk$^{1,2}$ }
\author{Fabien Bretenaker$^{1,2}$}
\email{Fabien.Bretenaker@u-psud.fr}
 \affiliation {$^1$Laboratoire Aim\'e Cotton, CNRS-ENS Cachan-Universit\'e Paris Sud 11, 91405 Orsay Cedex, France\\$^2$Ecole Polytechnique, 91128 Palaiseau Cedex, France}%Lines break automatically or can be forced with \\

\date{\today}% It is always \today, today,
             %  but any date may be explicitly specified

\begin{abstract}
The properties of a dual-frequency vertical external cavity surface emitting laser (VECSEL), in which two linear orthogonal polarization modes are oscillating simultaneously, are theoretically investigated. We derive a model based on the ideas introduced by San Miguel et al. [San Miguel, Feng, and Moloney, Phys. Rev. A \textbf{52}, 1728 (1995)], taking into account the spin dynamics of the carriers inside the quantum well (QW) based gain medium of the dual-frequency VECSEL. This model is shown to succeed in describing quite a few properties of the dual-frequency VECSEL, such as the behavior of nonlinear coupling strength between the modes, the spectral behavior of intensity noises of the modes, and also the correlation between these intensity noises. A good agreement is found with experimental data. The variables associated with the spin dependent carrier dynamics can be adiabatically eliminated due to the class-A dynamical behavior of the considered laser which is based on a cm-long external cavity. This leads to a simple analytical description of the dynamics of the dual-frequency VECSEL providing a better understanding of the physics involved.
%\begin{description}
%\item[Usage]
%Secondary publications and information retrieval purposes.
%\item[PACS numbers]
%May be entered using the \verb+\pacs{#1}+ command.
%\item[Structure]
%You may use the \texttt{description} environment to structure your abstract;
%use the optional argument of the \verb+\item+ command to give the category of each item. 
%\end{description}
\end{abstract}
\pacs{42.55.Ah, 42.55.Px, 42.60.Mi}

%\pacs{Valid PACS appear here}% PACS, the Physics and Astronomy
                             % Classification Scheme.
%\keywords{Suggested keywords}%Use showkeys class option if keyword
                              %display desired
\maketitle
\section{Introduction}
The spin-flip model \cite{SanMiguel1995}, abbreviated as SFM model, has been very successful in explaining different polarization properties of quantum well-based Vertical Cavity Surface Emitting Lasers (VCSELs), such as  selection of polarization state, polarization dynamics, switching of polarization state. This model considers two different electronic transitions associated to different magnetic sublevels of the conduction and heavy-hole valence bands and thus giving rise to gain for the two circularly polarized components of laser light of opposite helicities. Later, the SFM model has been extended to incorporate the effect of material birefringence (linear phase anisotropy), gain dichroism (gain anisotropy), saturable dispersion (linewidth enhancement factor or Henry factor), etc., which dictate different stability criteria of the linearly polarized modes of a VCSEL \cite{Travagnin1996, Martin-Regelado1996, Martin-Regelado1997, Travagnin1997, vanExter1998, Gahl1999, Prati2002}. Moreover, many other properties such as the effect of nonlinear phase anisotropy \cite{MPvanExter1998}, influence of polarization dynamics on the noise properties \cite{Mulet2001, Kaiser2002}, existence of elliptical or higher order transverse modes and possibilities of  having chaotic behavior even in free running lasers \cite{Virte2013}, have been quite successfully explained based on the fundamental ideas of spin dependent carrier dynamics as described in the SFM model. 

Recently, the possibility for Vertical External Cavity Surface Emitting Lasers (VECSELs) to sustain simultaneous oscillation of two orthogonal linearly polarized modes of slightly different frequencies has been evidenced \cite{Baili2009}. These dual-frequency lasers are very useful for microwave photonics applications \cite{Seeds2006} such as to generate high purity optically-carried tunable radio-frequency (RF) signals, which are essential for metrology, remote sensing, and analog communication applications \cite{Alouini2001,Tonda2006,Pillet2008,Camargo2012}.  Such dual-frequency oscillation has also been achieved in different kinds of solid-state lasers \cite{Brunel1997, Alouini1998, Czarny2004}. But the inherent class-B dynamical behavior of such solid-state lasers, as the population inversion lifetime is much longer than the intra-cavity photon lifetimes, gives rise to strong intensity noises resonant at the relaxation oscillation frequencies \cite{Arecchi1984, Taccheo1996}. This low-frequency intensity noise limits the spectral purity of the beatnote generated by dual-frequency solid-state lasers. On the contrary, the dual-frequency VECSEL overcomes this limitation since the cm-long external cavity ensures class-A dynamical behavior of the laser by making intra-cavity photon lifetime much longer than the carrier lifetime inside the semiconductor active medium \cite{Baili2009, De2013}. Moreover, to obtain stable simultaneous oscillation of the two modes, the nonlinear coupling constant between the modes is reduced below unity by spatially separating the two modes inside the gain medium with an intra-cavity birefringent crystal. The dependence of this nonlinear coupling constant on the spatial separation between the modes has already been  experimentally investigated  \cite{Pal2010}. Furthermore, the spectral behavior of the correlation between the intensity noises of the two modes has also been investigated experimentally and theoretically \cite{De2013}. These correlations play an important role to determine the purity of the RF beatnote since the intensity noise is coupled to the phase noise due to the large Henry factor of the quantum well gain medium \cite{Henry1982, Henry1986,De2014}. As described in Refs. \cite{De2013, Pal2010,De2014}, we developed a fully heuristic rate equation model for the laser, which completely overlooked the spin-dependent dynamics of the carriers as mentioned in the SFM model. This model was a standard two-mode rate equation model in which nonlinear cross-saturation coupling terms were added to mimic the nonlinear coupling between modes \cite{De2013}. Although this model was surprisingly successful to explain the behavior of nonlinear coupling and noise properties of the two linear polarized modes in the dual-frequency VECSEL, it seems important to be able to justify it on more solid grounds, taking into account in particular the spin-dependent dynamics in the quantum wells. 

The purpose of the present paper is thus to analyze the properties of the two orthogonally polarized modes of a dual-frequency VECSEL from the perspective of the SFM model. We start by writing the rate equations for our laser based on the ideas of the SFM model and taking into account the partial spatial overlap between the modes. We also aim at adiabatically eliminating the carrier dynamics as our laser is class-A type, hoping to be able then to derive the heuristic equations that we have used till now \cite{De2013, De2014}. 

The paper is organized as follows. In Sec. II, we present the rate equation model relevant for our dual-frequency VECSEL and based on the ideas of the SFM model. Section III describes how these rate equations can be simplified thanks to adiabatic elimination of the fast variables and averaging over rapidly varying phases. In Sec. IV we use this model to describe the nonlinear coupling between the two linear eigenpolarizations. The intensity noise properties are investigated in Sec. V. Finally, Sec. VI summarizes the main conclusions of this paper.    

\section{Modeling the dual-frequency VECSEL}
Before developing the model for the dual-frequency VECSEL, we first remind the standard SFM model in the first subsection of this section. This model will be generalized to the dual-frequency VECSEL in the next subsection.
\subsection{SFM VCSEL model in linear polarization basis}
The standard SFM model, which has been primarily developed to explain the polarization properties of VCSELs, considers the coexistence of two different electronic transitions providing gain to the two circularly polarized components of light of opposite helicities \cite{SanMiguel1995, Travagnin1996}. 

\begin{figure}[]
\includegraphics[width=0.4\textwidth]{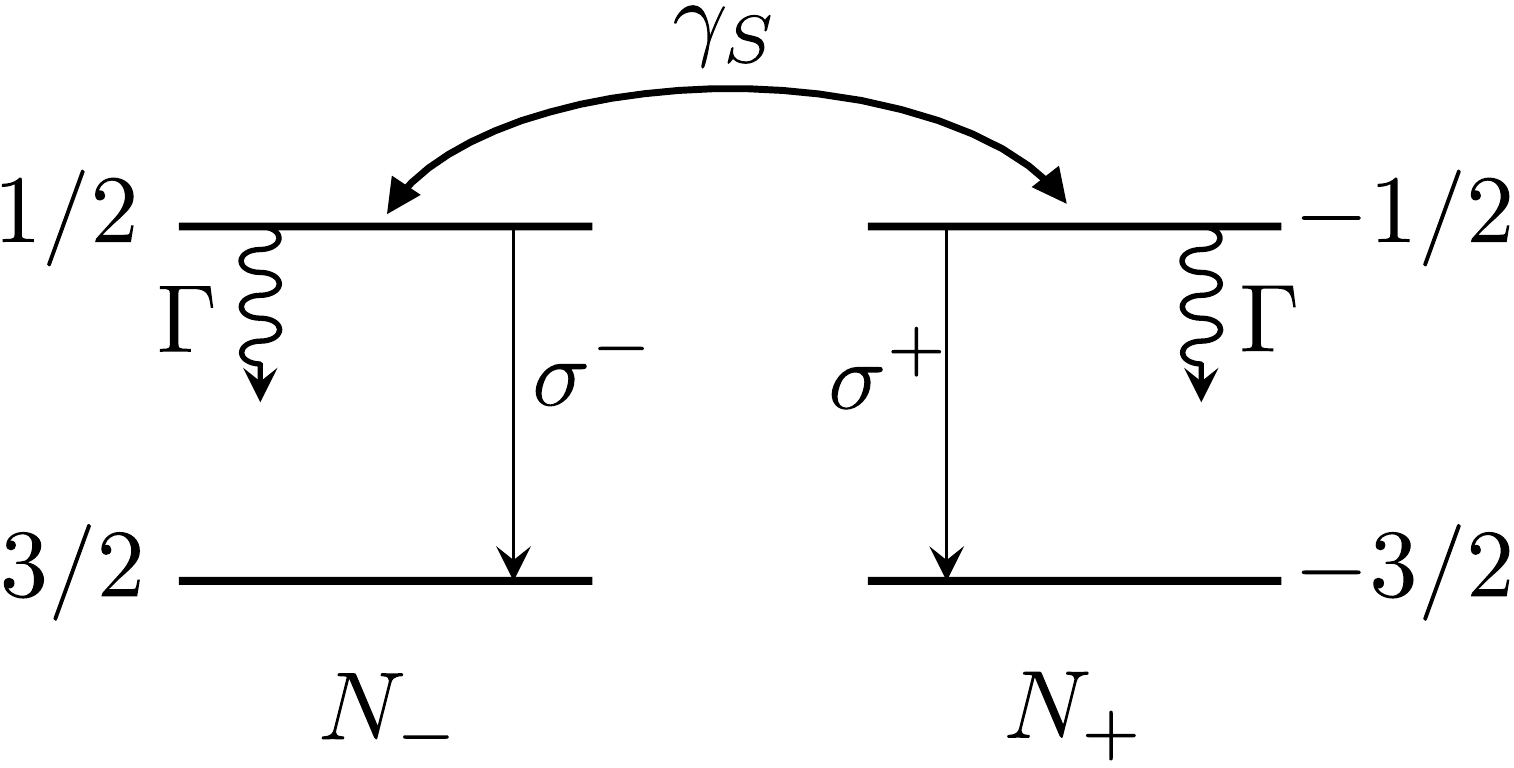}
\caption{ Level scheme involved in the SFM model \cite{SanMiguel1995}. $N_+$ and $N_-$ are the population inversions for the $\sigma^+$ and $\sigma^-$ transition, respectively. The decay rate of these population inversions is $\Gamma$. The spin-flip rate is $\gamma_S$.}\label{Fig00}
\end{figure}

In the basis of linear eigenpolarizations the dynamics of the standard VCSEL can be described by the following rate equations \cite{Martin-Regelado1996, Martin-Regelado1997, Travagnin1997}:
\begin{eqnarray}
\dfrac{dE_{x}}{dt}&=&-\dfrac{\gamma_{x}}{2} E_{x}-i (\kappa \alpha + \gamma_{p}) E_{x}\nonumber\\
 &&+ \dfrac{\kappa}{2} (1+i\alpha)(NE_{x} + inE_{y}) ,\label{Eq01}\\
\dfrac{dE_{y}}{dt}&=&-\dfrac{\gamma_{y}}{2} E_{y}-i (\kappa \alpha - \gamma_{p}) E_{y}
\nonumber\\
 &&+ \dfrac{\kappa}{2} (1+i\alpha)(NE_{y} - inE_{x}) ,\label{Eq02} \\
\dfrac{dN}{dt}&=&-\Gamma(N-N_{0})-\kappa N(\vert E_{x} \vert ^{2} + \vert E_{y} \vert ^{2})
\nonumber\\ 
&&- i\kappa n( E^*_{x}E_{y} - E_{x}E^*_{y}) ,\label{Eq03}\\
\dfrac{dn}{dt}&=&-\gamma_{S}n-\kappa n(\vert E_{x} \vert ^{2} + \vert E_{y} \vert ^{2}) \nonumber\\
&&- i\kappa N( E^*_{x}E_{y} - E_{x}E^*_{y}) .\label{Eq04}
\end{eqnarray}
In these equations, $ E_x $ and $ E_y $ are the dimensionless slowly varying complex amplitudes of the two orthogonal linearly polarized fields. $ N=N_+ + N_- $ is the total population inversion, where $N_+$ (resp. $N_-$) is the population inversion for the transition amplifying $\sigma^+$- (resp. $\sigma_-$-) polarized light (see Fig.\ \ref{Fig00}). $ n=N_+ - N_-$ is the population inversion difference between the transitions between the magnetic sublevels, i. e., the difference between the population inversions for the $\sigma^+$ and $\sigma^-$ transitions ; $ N_0 $ defines the unsaturated population inversion, which is proportional to the pumping rate. $N$, $n$, and $N_0$ are dimensionless quantities, i. e., numbers of atoms or differences between numbers of atoms in the different levels. $ \gamma_x$ and $\gamma_y$ are the intensity decay rates for the two polarization states inside the cavity. $ \gamma_x\neq\gamma_y$ takes into account the possible linear dichroism  of the cavity. $ \gamma_{p} $ holds for the linear phase anisotropy of the cavity. $ \Gamma $ represents the decay rate of total population inversion, whereas $ \gamma_S $ represents the decay rate of the population inversion difference (spin-flip rate). The stimulated emission coefficient $ \kappa $ is proportional to the stimulated emission cross-section and $ \alpha $ holds for the Henry phase-amplitude coupling factor \cite{Henry1982, Henry1986}. 

\subsection{Extension to the dual-frequency VECSEL with partial spatial separation}
\begin{figure}[]
\includegraphics[width=0.4\textwidth]{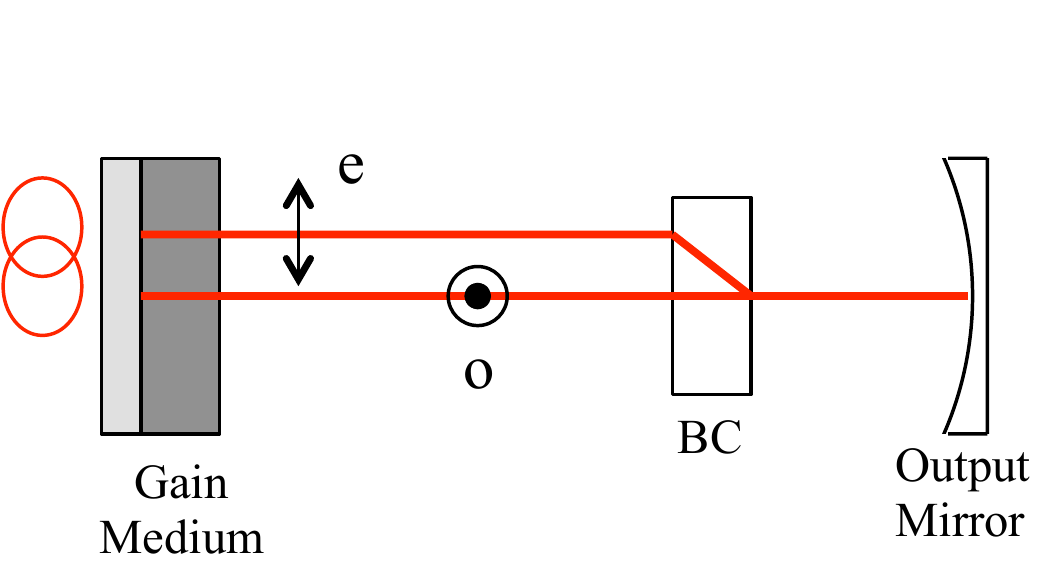}
\caption{ (Color online) Schematic representation of the dual-frequency VECSEL. The intra-cavity birefringent crystal (BC) spatially separates the ordinary (o) and extraordinary (e) polarized modes inside the gain medium. }\label{Fig01}
\end{figure}\begin{figure}[]
\includegraphics[width=0.4\textwidth]{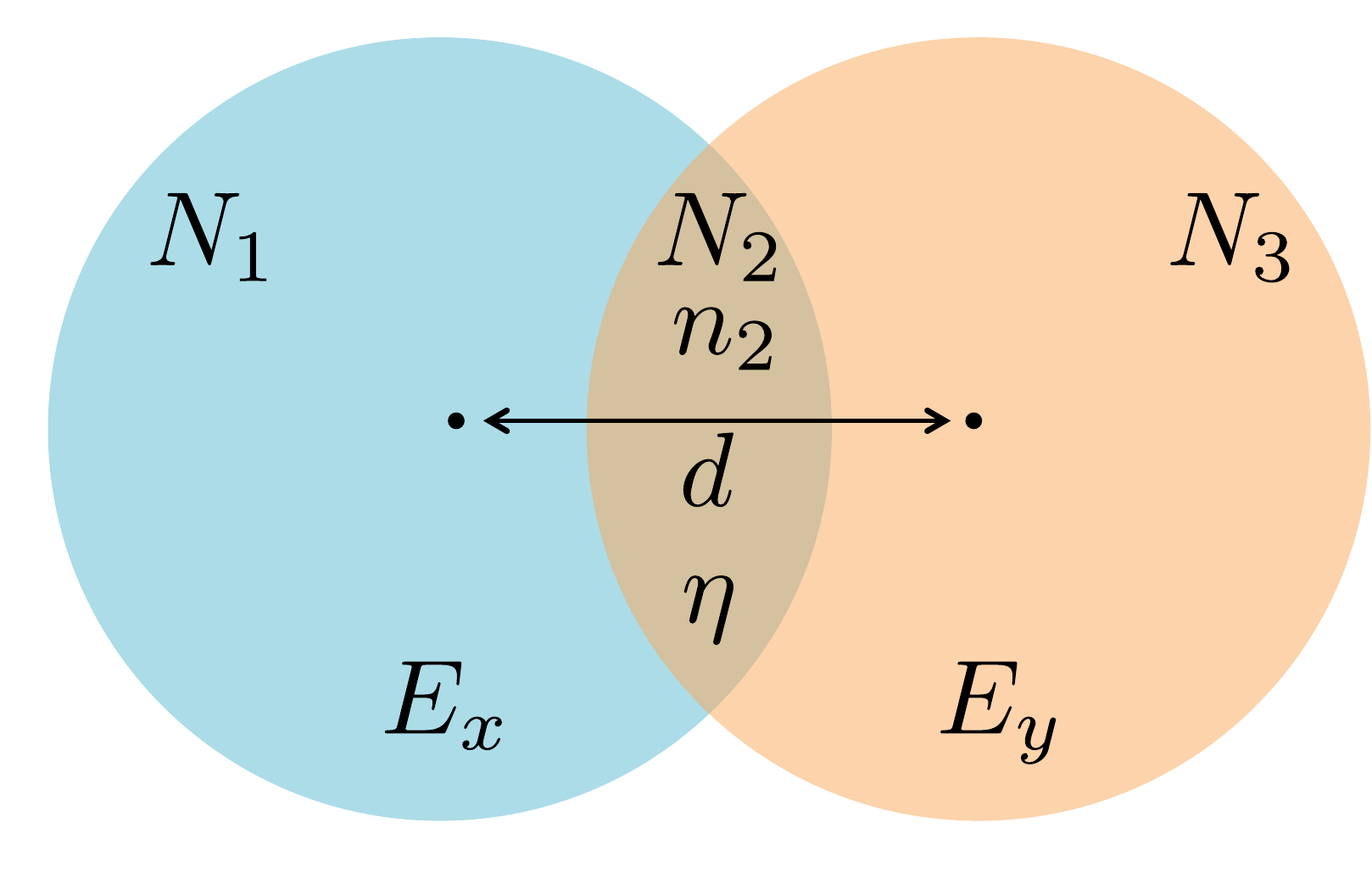}
\caption{ (Color online) Schematic representation of the two modes in the gain medium of the dual-frequency VECSEL. We model the two beams as two top-hat cylindrical beams with circular sections. $ E_{x} $ (resp. $E_y$) is the complex amplitude of the $x$-polarized (resp. $y$-polarized) field. $N_1$ (resp. $N_3$) is the total population inversion in the region where only the $x$-polarized (resp. $y$-polarized) mode has a non-zero intensity. $N_2$ is the total population inversion and $n_2$ is the population inversion difference in the region where both modes are superimposed. The parameters $ d $ and $ \eta $ respectively represent the spatial separation and the relative overlap between the two modes.}\label{Fig02}
\end{figure}
Our aim here is to model the dual-frequency VECSEL, sustaining the oscillation of two orthogonal linearly polarized modes, which are partially spatially separated by an intra-cavity birefringent crystal (BC) as schematized in Fig.\ \ref{Fig01}. 
This dual-frequency VECSEL has two main differences compared to the standard VCSEL. The first difference is the cm-long external cavity, which ensures class-A dynamical behavior by increasing the intra-cavity photon lifetime ($ \sim 5-10\ \mathrm{ns}$) compared to the population inversion lifetime ($ \sim 3\ \mathrm{ns}$) \cite{Baili2009,  Pal2010,De2013,Baili2009b}, whereas in standard VCSELs the photon lifetime ($ \sim 1\ \mathrm{ps}$) inside the cavity is usually much shorter than the population inversion lifetime ($ \sim 1\ \mathrm{ns}$), leading to class-B dynamical behavior \cite{Scott1994, Tauber1993}.  The second difference comes from the intra-cavity BC, which spatially separates the two modes inside the gain structure. This ensures simultaneous oscillation of the two orthogonally polarized modes by reducing their nonlinear coupling. This nonlinear coupling is lowered as the partial overlap between the modes reduces the cross-saturation effect. Moreover, varying spatial overlap between the modes permits to vary the strength of the nonlinear coupling \cite{Pal2010}. Besides, the intra-cavity BC introduces a strong linear phase anisotropy inside the cavity, which creates a large frequency difference (few GHz) between the two modes which are polarized along the ordinary and extraordinary eigenpolarizations of the BC \cite{Baili2009}. Therefore, the main difficulty to describe our dual-frequency VECSEL in the spirit of the SFM model is to introduce the effect of the partial spatial overlap between the modes. This is taken care of by introducing the dimensionless parameter $ \eta $ ($0\leq\eta\leq1$), which we define as: 
\begin{eqnarray}
\eta &=& \dfrac{\int I(x,y)I(x-d,y) dxdy }{\sqrt{\int I^2(x,y)dxdy \int I^2(x-d,y)dxdy}}\nonumber\\
&=&\exp{\left(-\dfrac{d^2}{w^2}\right)}\ ,
\end{eqnarray}
where $ I(x,y) $  denotes the Gaussian intensity profiles of radius $w$ of the laser mode in the active medium and  $ d $ is the spatial separation between the axes of the ordinary and extraordinary modes inside the gain structure (see Fig.\ \ref{Fig02}). Of course this parameter $\eta$ could also be used to model other effects which tend to increase the effective overlap between the two modes, such as carrier diffusion, etc.

The basic hypotheses of our model are schematized in Fig.\ \ref{Fig02}. For the sake of simplicity, we model the two modes as two top-hat cylindrical beams with circular sections, with an overlap $\eta$.  We then choose to consider three population reservoirs associated with the three different regions schematized in Fig.\ \ref{Fig02}: Region 1 (resp. 3) is the region where only the $x$-polarized (resp. $y$-polarized) mode is present with a complex field amplitude $E_x$ (resp. $E_y$) while region 2 is the place where both modes overlap. We then model the dual-frequency VECSEL using the following rate equations: 
\begin{eqnarray}
\dfrac{dE_{x}}{dt}&=&-\dfrac{\gamma_{x}}{2}E_{x}-i(\kappa \alpha + \gamma_{p})E_{x}
\nonumber\\
&&+\dfrac{\kappa}{2}(1+i\alpha)\left[(N_{1}+N_2)E_{x}+in_2 E_{y}\right]\ ,\label{Eq06}\\
\dfrac{dE_{y}}{dt}&=&-\dfrac{\gamma_{y}}{2}E_{y}-i(\kappa \alpha - \gamma_{p})E_{y}
\nonumber\\
&&+\dfrac{\kappa}{2}(1+i\alpha)\left[(N_{2}+N_3)E_{y}-in_2 E_{x}\right]\ ,\label{Eq07}\\
\dfrac{dN_{1}}{dt}&=&-\Gamma (N_{1} - N_{01}) - \kappa N_{1}\vert E_{x} \vert ^2 ,\label{Eq08}\\
\dfrac{dN_{2}}{dt}&=&-\Gamma( N_{2} - N_{02}) - \kappa N_{2}(\vert E_{x} \vert ^2 + \vert E_{y} \vert ^2)
\nonumber\\
&& - i\kappa n_2 (E^*_{x}E_{y}-E_{x}E_{y}^*)\ ,\label{Eq09}\\
\dfrac{dN_{3}}{dt}&=&-\Gamma (N_{3} - N_{03}) - \kappa N_{3}\vert E_{y} \vert ^2 ,\label{Eq10} \\
\dfrac{dn_2}{dt}&=&-\gamma_{S} n_2 - \kappa n_2(\vert E_{x} \vert ^2 + \vert E_{y} \vert ^2)
\nonumber\\
&& - i\kappa N_2(E^*_{x}E_{y}-E_{x}E_{y}^*)\ ,\label{Eq10N1}
\end{eqnarray}
where the notations $ E_{x} $, $ E_{y} $, $ \gamma _x$, $\gamma_y$, $ \Gamma $, $ \gamma_{S} $, $ \gamma_{p} $, $ \kappa $, and $ \alpha $ have the same meaning as in Eqs. (\ref{Eq01}-\ref{Eq04}).
 It is worth noticing that all populations entering these equations, namely $N_1$, $N_2$, $N_3$, and $n_2$, and their pumping terms $N_{01}$, $N_{02}$, and $N_{03}$, are dimensionless number of atoms. Since only the $x$-polarized (resp. $y$-polarized) mode is present in region 1 (resp. 2) of Fig.\ \ref{Fig02}, we need to introduce only a total population inversion $N_1$ (resp. $N_3$) in this region, and no population inversion difference $n_1$ (resp. $n_3$) is created in this region. On the contrary, in region 2 where the two modes overlap, we introduce both the total population inversion $N_2$ and the population inversion difference $n_2$.  
 
Since the quantities $N_1$, $N_2$, and $N_3$ that we use in Eqs.\,(\ref{Eq06}-\ref{Eq10N1}) are dimensionless number of atoms, they are already the results of the integration of the density of atoms over the transverse mode field distributions. They consequently already take into account the confinement factors of the two modes that appears explicitly when the equations are written in terms of atomic densities \cite{Lamb1974,Wieczorek2004,Wieczorek2005}. Here, the overlapping factor $\eta$ between the two modes enters the model through the pumping rates $N_{01}$, $N_{02}$, and $N_{03}$. More precisely, the amount of overlap determines the strength of the pumping $N_{02}$ that goes to the region of the active medium where the two modes overlap with respect to the pumping rates  $N_{01}$ and $N_{03}$ in the regions where only one of the modes has a non vanishing intensity. Quantitatively, if we introduce the total pumping rates $N_{0x}$ and $N_{0y}$ for the $x$- and $y$-polarized modes, respectively, then $N_{01}$, $N_{02}$, and $N_{03}$ are defined by:
\begin{eqnarray}
N_{01}&=&N_{0x} -\eta \frac{N_{0x}+N_{0y}}{2} ,\label{Eq10N2}\\
N_{02}&=&\eta \frac{N_{0x}+N_{0y}}{2} ,\label{Eq10N3}\\
N_{03}&=&N_{0y}-\eta \frac{N_{0x}+N_{0y}}{2} .\label{Eq10N4}
\end{eqnarray}

One can notice that if we suppose that one of the modes vanishes (for example $E_y=0$ or $N_{02}=N_{03}=0$), we then retrieve the two usual rate equations for $E_x$ and $N_{x}=N_1+N_2$. Moreover, in the case where the modes totally overlap ($\eta=1$ and $N_{0x}=N_{0y}$, or similarly $N_{01}=N_{03}=0$ and $N_{02}\neq0$), Eqs.\ (\ref{Eq06}-\ref{Eq10N1}) reduce to Eqs.\ (\ref{Eq01}-\ref{Eq04}) with $N=N_2$, $n=n_2$, and $N_1=N_3=0$. 

\section{Simplification of the rate equations}
\subsection{Adiabatic elimination of the spin dynamics}

The parameter $ \gamma_{S} $ takes into account various microscopic spin-flip relaxation processes, which tend to equilibrate the populations of the magnetic sublevels inside the quantum-well-based active medium. Measurements have shown that $\gamma_{S}^{-1} $ is of the order of a few tens of picoseconds inside quantum wells \cite{Damen1991, Bar-Ad1992}, whereas $ \Gamma^{-1}\approx 3\;\mathrm{ns} $ and $ \gamma_x^{-1},\gamma_y^{-1}\approx 5-10\;\mathrm{ns} $ in the structures used to build the present dual-frequency VECSEL, as shown by previous experiments performed with the same 1/2-VCSEL \cite{Baili2009,Pal2010, De2013}.  Therefore, we can adiabatically eliminate the dynamics of $ n_2$ from the rate equations by taking $dn_2/dt=0$ in Eq.\ (\ref{Eq10N1}). This leads to :
\begin{equation}
n_2= \dfrac{-i\kappa N_2(E^*_{x}E_{y}-E_{x}E^*_{y})}{\gamma_{S}+\kappa (\vert E_{x} \vert ^2 + \vert E_{y} \vert ^2)}\ . \label{Eq12}
\end{equation}
To obtain Eq.\ (\ref{Eq12}), we have also taken into account the fact that the relative phase between the two fields evolves slower than the spin relaxation rate, namely $\gamma_p \ll \gamma_S$.

We can now suppose that the field amplitudes $E_x$ and $E_y$ in Eqs.\ (\ref{Eq06}-\ref{Eq10N1}) are normalized in such a way that the squares of their moduli correspond to the numbers of photons $F_x$ and $F_y$ inside the cavity for the two polarization states, leading to: 
\begin{equation}
E_{x}=\sqrt{F_{x}}e^{i\phi_{x}}\ ,\ E_{y}=\sqrt{F_{y}}e^{i\phi_{y}} \ ,\label{Eq14}
\end{equation} 
where we have introduced the phases $\phi_x$ and $\phi_y$ of the two modes. We define their difference as
\begin{equation}
\phi=\phi_{x}-\phi_{y} .\label{Eq15}
\end{equation}
 With these notations, Eq. (\ref{Eq12}) reads 
\begin{equation}
n_2=-2\varepsilon \dfrac{\kappa}{\Gamma}N_2\sqrt{F_{x}F_{y}}\sin\phi\ ,\label{Eq16}
\end{equation}
where we have supposed that the large value of $\gamma_S$ leads to $\displaystyle{ \frac{\kappa}{\gamma_S}F_{x,y}\ll 1 }$ and where we have introduced 
\begin{equation} \varepsilon = \Gamma/\gamma_{S} ,
\end{equation}
 which is also supposed to be much smaller than 1. Introducing this steady state-values of $ n $ into Eqs. (\ref{Eq06}-\ref{Eq10}) and using the notations defined in Eqs. (\ref{Eq14},\ref{Eq15}), we obtain :
%\begin{widetext}
\begin{eqnarray}
\dfrac{dF_{x}}{dt}&=&-\gamma_{x}F_{x}+\kappa(N_1+N_2) F_{x}\nonumber\\
&&-2\kappa^2\frac{\varepsilon}{\Gamma}N_2F_xF_y\sin\phi(\sin\phi-\alpha\cos\phi)\  ,\label{Eq19}\\
\dfrac{dF_{y}}{dt}&=&-\gamma_{y}F_{y}+\kappa(N_3+N_2) F_{y}\nonumber\\
&&-2\kappa^2\frac{\varepsilon}{\Gamma}N_2F_xF_y\sin\phi(\sin\phi+\alpha\cos\phi)\ ,\label{Eq20}\\
\dfrac{dN_{1}}{dt}&=&-\Gamma(N_{1}-N_{01})-\kappa N_{1}F_{x}\ ,\label{Eq21}\\
\dfrac{dN_{2}}{dt}&=&-\Gamma(N_{2}-N_{02})-\kappa N_{2}(F_{x}+F_{y})\nonumber\\
&&+2\kappa^2\frac{\varepsilon}{\Gamma}N_2F_xF_y\ ,\label{Eq22}\\
\dfrac{dN_{3}}{dt}&=&-\Gamma(N_{3}-N_{03})-\kappa N_{3}F_{y}\ ,\label{Eq23}\\
\dfrac{d\phi}{dt}&=&-2\gamma_{p}+\dfrac{\kappa}{2}\alpha(N_{1}-N_{3})\nonumber\\
&&-\dfrac{\kappa^2\varepsilon}{2\Gamma}N_2
\left[F_{y}(\sin 2 \phi+2\alpha\sin^2\phi)\right.\nonumber\\
&&\left.-F_{x}(\sin 2 \phi-2\alpha\sin^2\phi)\right]\ . \label{Eq23N1}
\end{eqnarray}
%\end{widetext}

\subsection{Averaging the relative phase}
Previous experiments have shown that the frequency difference between the two orthogonally polarized modes, which is ruled mainly by the term $ 2\gamma_{p}$ coming from the intracavity phase anisotropy in Eq.\ (\ref{Eq23N1}), is of the order of a few gigahertz \cite{Baili2009}. This is the range of frequencies which is also interesting for microwave photonics applications. Such frequencies are much larger than the population inversion decay rate ($ \Gamma \approx 3\times 10^8\; \mathrm{s}^{-1} $) in the semiconductor structures we consider here \cite{Baili2009b}. In these conditions, this permits us to further simplify  the rate equations (\ref{Eq19}-\ref{Eq23}) by averaging over all values $ \phi $ from 0 to $ 2\pi $. This leads to 
 \begin{eqnarray}
\dfrac{dF_{x}}{dt}&=&-\gamma_{x}F_{x}+\kappa(N_1+N_2) F_{x}-\kappa^2\frac{\varepsilon}{\Gamma}N_2F_xF_y ,\label{Eq24}\\
\dfrac{dF_{y}}{dt}&=&-\gamma_{y}F_{y}+\kappa(N_2+N_3) F_{y}-\kappa^2\frac{\varepsilon}{\Gamma}N_2F_xF_y ,\label{Eq25}\\
\dfrac{dN_{1}}{dt}&=&-\Gamma(N_{1}-N_{01})-\kappa N_{1}F_{x}\ ,\label{Eq26}\\
\dfrac{dN_{2}}{dt}&=&-\Gamma(N_{2}-N_{02})-\kappa N_{2}(F_{x}+F_{y})\nonumber\\
&&+2\kappa^2\frac{\varepsilon}{\Gamma}N_2F_xF_y\ ,\label{Eq27}\\
\dfrac{dN_{3}}{dt}&=&-\Gamma(N_{3}-N_{03})-\kappa N_{3}F_{y}\ .\label{Eq27N0}
\end{eqnarray}

\subsection{Class-A approximation}
 
The cm-long external cavity of our dual-frequency VECSEL ensures a longer lifetime for the intra-cavity photons than for the population inversion ($\gamma_x,\gamma_y\ll\Gamma$). Indeed, the results of previous experiments have proved that the photon lifetimes ($ \gamma_{x}^{-1} $,$ \gamma_{y}^{-1} $) are about 10 ns, whereas the population inversion lifetime ($ \Gamma^{-1} $) is about 3 ns \cite{Baili2009, Pal2010, De2013}.  The dual-frequency VECSEL is thus a class-A laser. Therefore, we can adiabatically eliminate $N_{1}$, $N_2$, and $ N_{3} $ from Eqs. (\ref{Eq24}-\ref{Eq27N0}) by writing $dN_{1}/dt=0$, $dN_{2}/dt=0$, and $dN_{3}/dt=0$ in Eqs. (\ref{Eq26}-\ref{Eq27N0}). 
This leads to:
 \begin{eqnarray}
N_{1}&=&\dfrac{N_{01}}{1+\frac{\kappa}{\Gamma}F_x}\ ,\label{Eq27N1}\\
N_{2}&=&\dfrac{N_{02}}{1+\frac{\kappa}{\Gamma}(F_x+F_y)-2\frac{\kappa^2\varepsilon}{\Gamma^2}F_xF_y}\ ,\label{Eq27N2}\\
N_{3}&=&\dfrac{N_{03}}{1+\frac{\kappa}{\Gamma}F_y}\ .\label{Eq27N2bis}
\end{eqnarray}
By injecting Eqs.\ (\ref{Eq27N1}-\ref{Eq27N2bis}) into Eqs.\ (\ref{Eq24}) and (\ref{Eq25}), we are left with the following differential equations: 
\begin{widetext}
\begin{eqnarray}
\dfrac{dF_{x}}{dt}&=&-\gamma_{x}F_{x}+\kappa\left[\dfrac{N_{01}}{1+\frac{\kappa}{\Gamma}F_x}+\dfrac{N_{02}}{1+\frac{\kappa}{\Gamma}(F_x+F_y)-2\frac{\kappa^2\varepsilon}{\Gamma^2}F_xF_y}\left(1-\frac{\kappa\varepsilon}{\Gamma}F_y\right)\right]F_x\ , \label{Eq27N3}\\
\dfrac{dF_{y}}{dt}&=&-\gamma_{y}F_{y}+\kappa\left[\dfrac{N_{03}}{1+\frac{\kappa}{\Gamma}F_y}+\dfrac{N_{02}}{1+\frac{\kappa}{\Gamma}(F_x+F_y)-2\frac{\kappa^2\varepsilon}{\Gamma^2}F_xF_y}\left(1-\frac{\kappa\varepsilon}{\Gamma}F_x\right)\right]F_y\ . \label{Eq27N4}
\end{eqnarray}
\end{widetext}
\begin{figure*}[]
\includegraphics[width=1.0\textwidth]{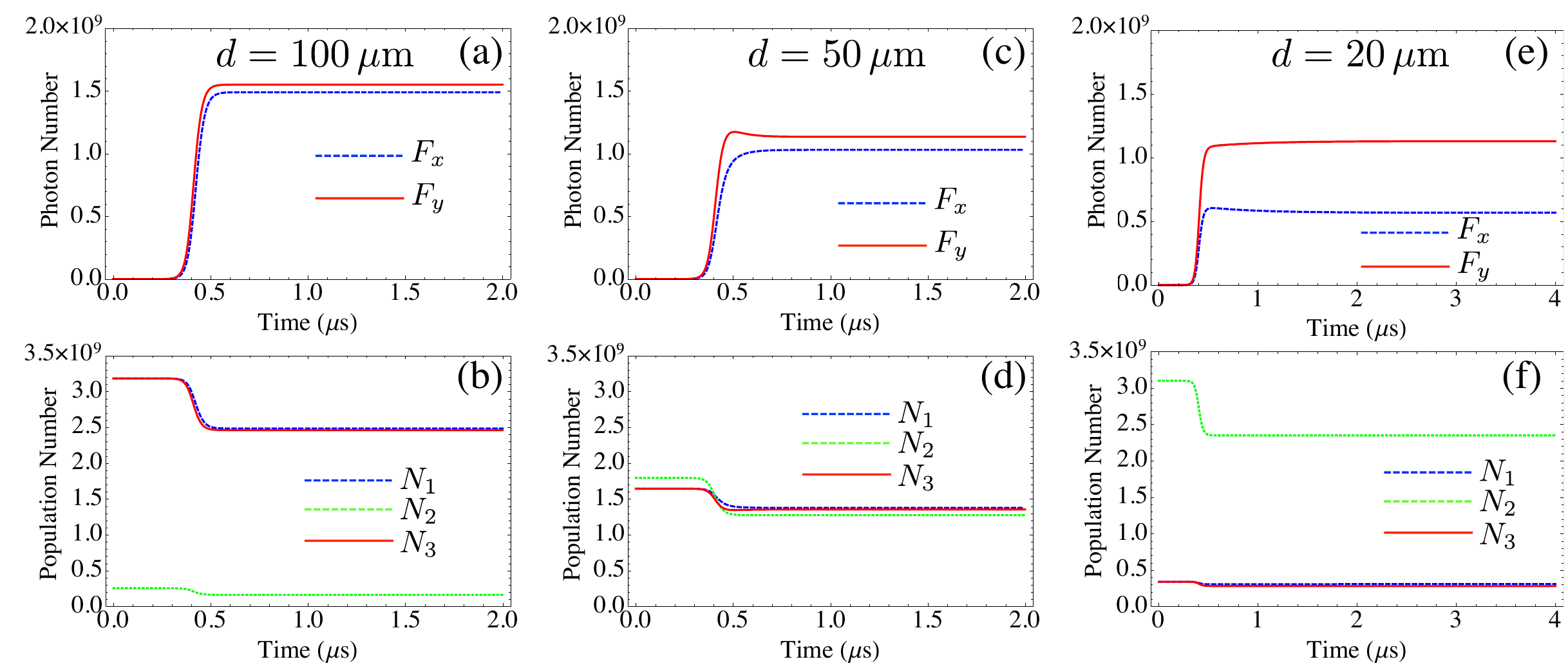}
\caption{(Color online) Time evolution of (a) the intra-cavity photon numbers $ F_{x} $ and $ F_{y} $ and (b) the population inversions $N_1$, $N_2$, and $N_3$ for $d=100\,\mu\mathrm{m}$, $w=62\,\mu\mathrm{m}$, $1/\gamma_x=6\,\mathrm{ns}$, $1/\gamma_y=6.05\,\mathrm{ns}$, $1/\Gamma=3\,\mathrm{ns}$, $\varepsilon=0.02$, $\kappa=6.3\times 10^{-2}\,\mathrm{s}^{-1}$, $N_{0x}=N_{0y}=1.3\times\gamma_x/\kappa$. In this case $\eta=0.074$. (c),(d): Same as (a),(b) for $d=50\,\mu\mathrm{m}$, corresponding to $\eta=0.52$. (e),(f): Same as (a),(b) for $d=20\,\mu\mathrm{m}$, corresponding to $\eta=0.90$.}\label{Fig04}
\end{figure*}

\subsection{Weak saturation limit}
In the case where the laser is just above threshold and the saturation of the active medium is weak, i. e.,   $ \kappa F_{x}/\Gamma,\kappa F_{y}/\Gamma\ll 1 $, one can keep only terms up to first order in $ \kappa F_{x}/\Gamma$ and $\kappa F_{y}/\Gamma $ in Eqs. (\ref{Eq27N3}) and (\ref{Eq27N4}), leading to the following ``third order in field'' version for Eqs. (\ref{Eq27N3}) and (\ref{Eq27N4}): 
\begin{eqnarray}
\dfrac{dF_{x}}{dt}&=&-\gamma_{x}F_{x}+\kappa \left[(N_{01}+N_{02})(1-\frac{\kappa}{\Gamma}F_x)\right.\nonumber\\
&&\left.-\frac{\kappa}{\Gamma}N_{02}(1+\varepsilon)F_y\right]F_x\;,\label{Eq30}\\
\dfrac{dF_{y}}{dt}&=&-\gamma_{y}F_{y}+\kappa \left[(N_{03}+N_{02})(1-\frac{\kappa}{\Gamma}F_y)\right.\nonumber\\
&&\left.-\frac{\kappa}{\Gamma}N_{02}(1+\varepsilon)F_x\right]F_y\;.\label{Eq31}
\end{eqnarray}
Equations (\ref{Eq30}) and (\ref{Eq31}) end up in the same form as the classical equations of Lamb's theory \cite{Lamb1974}. In particular, the coefficients $ \xi_{xy} $, $ \xi_{yx} $ which define the ratios between the cross- to self-saturation coefficients and which describe the nonlinear coupling of the two laser modes are given by 
\begin{eqnarray}
 \xi_{xy}&=&\frac{N_{02}(1+\varepsilon)}{N_{01}+N_{02}}=\frac{\eta}{2}(1 + \varepsilon)\left(1 +\dfrac{N_{0y}}{N_{0x}}\right)\ ,\label{Eq32}\\
 \xi_{yx}&=&\frac{N_{02}(1+\varepsilon)}{N_{03}+N_{02}}=\frac{\eta}{2}(1 + \varepsilon)\left(1 +\dfrac{N_{0x}}{N_{0y}}\right)\ .\label{Eq33}
\end{eqnarray}
Equations \ (\ref{Eq30}) and (\ref{Eq31}) allow us to retrieve the heuristic model that was used in Refs. \cite{Pal2010, De2013}. But the model of Refs. \cite{Pal2010, De2013} did not allow to derive any expressions for $ \xi_{xy} $ and $ \xi_{yx} $, whereas the present model provides analytical expressions for these coupling coefficients [see Eqs.\ (\ref{Eq32}) and (\ref{Eq33})], showing their dependence on the amount  $ \eta $ of spatial overlap between the modes and the value of the spin-relaxation decay rate ($ \gamma_{S} $) relatively to the population inversion decay rate ($ \Gamma $).

\subsection{Laser switch-on}
We now use Eqs.\ (\ref{Eq24}-\ref{Eq27N0}) to simulate the switching on of the laser. The temporal evolutions of the intra-cavity photon numbers ($ F_{x} $, $ F_{y} $) of the two modes for different values of the spatial separation $ d $ are displayed in Fig.\;\ref{Fig04}. These values of $d$ correspond to those used in the experiments \cite{Pal2010, De2013}. To perform the numerical integration of Eqs.\ (\ref{Eq24}-\ref{Eq27N0}), we chose values of the parameters $w$, $\kappa$, $\Gamma$, $\gamma_x$, $\gamma_y$ representative of the experiment described in Ref.\ \cite{De2013}. Only the value of $\varepsilon$ is estimated from the orders of magnitude of $\gamma_S$ that one can find on the literature \cite{Damen1991,Bar-Ad1992}. We suppose that the laser is 1.3 times above threshold, that the two polarization eigenstates are equally pumped, i. e., $N_{0x}=N_{0y}$, but that the $x$-polarized mode experiences slightly more losses than the $y$-polarized one. The results of Fig. \ref{Fig04} show that the two polarization eigenstates can oscillate simultaneously for all three values of $d$, but that when $d$ decreases, the increase of the competition makes the final unbalance between their intensities more important. This is consistent with the picture expected from an increase of nonlinear coupling due to an increase of $\eta$. The comparison of the evolutions of the populations $N_1$, $N_2$, and $N_3$ in the three cases of Figs.\ \ref{Fig04}(b), \ref{Fig04}(d), and \ref{Fig04}(f) is also interesting. One can notice that when $d$ decreases, the increase of the overlap of the two modes leads to an increase of the relative weight of $N_2$ with respect to $N_1$ and $N_3$.  Moreover, this decrease of $d$ also leads to a decrease of the total number $N_1+N_2+N_3$ of atoms providing gain to the laser. This is consistent with an increase of the competition between the two modes.

\begin{figure}[h!]
\includegraphics[width=0.4\textwidth]{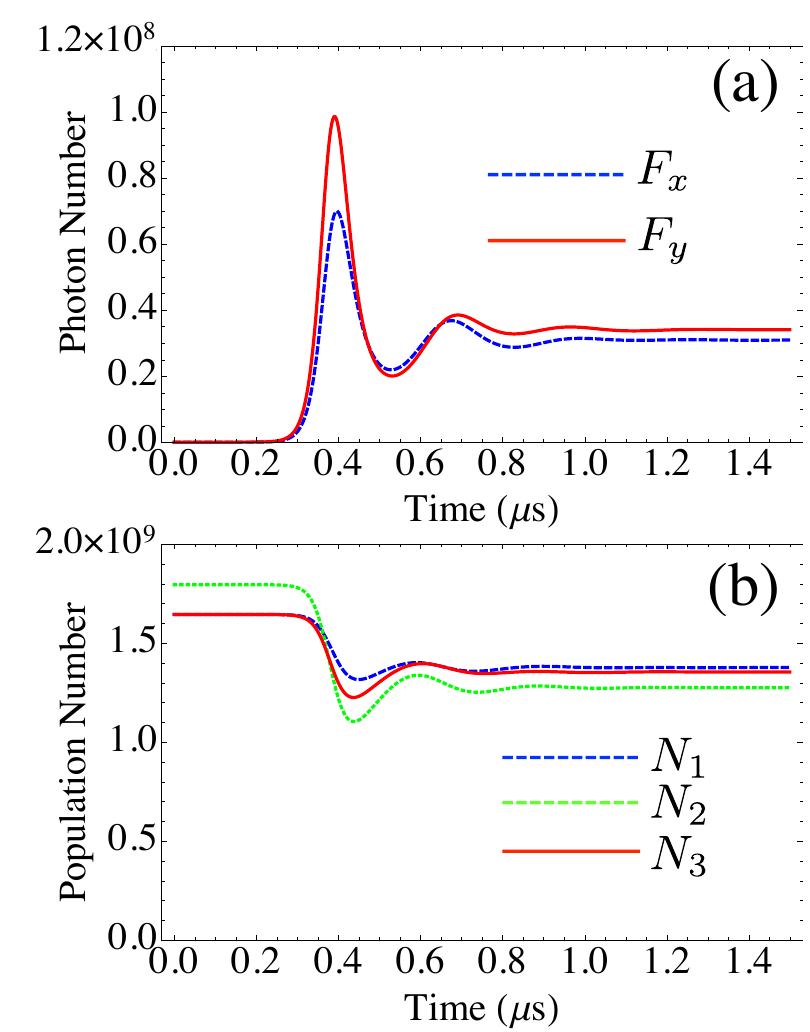}
\caption{(Color online) Same as Fig.\,\ref{Fig04}(c,d) in the case of a class-B lasers. The values of the parameters are $d=50\,\mu\mathrm{m}$, $w=62\,\mu\mathrm{m}$, $1/\gamma_x=6\,\mathrm{ns}$, $1/\gamma_y=6.05\,\mathrm{ns}$, $1/\Gamma=100\,\mathrm{ns}$, $\varepsilon=0.02$, $\kappa=6.3\times 10^{-2}\,\mathrm{s}^{-1}$, $N_{0x}=N_{0y}=1.3\times\gamma_x/\kappa$. }\label{Fig05}
\end{figure}

Before going further into the exploration of the behavior of steady-state solutions in our model, we check its consistency by verifying that it can also describe the well known damped relaxation behavior of a class-B laser. We thus strongly increase the value of $1/\Gamma$ and simulate the switching on of the laser using Eqs.\ (\ref{Eq24}-\ref{Eq27}). The result is reproduced in Fig.\ \ref{Fig05}. The laser clearly exhibits the in-phase and anti-phase relaxation oscillations, showing that our model consistently describes the behavior of class-B lasers also.

\subsection{Steady-state solutions}
\begin{figure}[]
\includegraphics[width=0.4\textwidth]{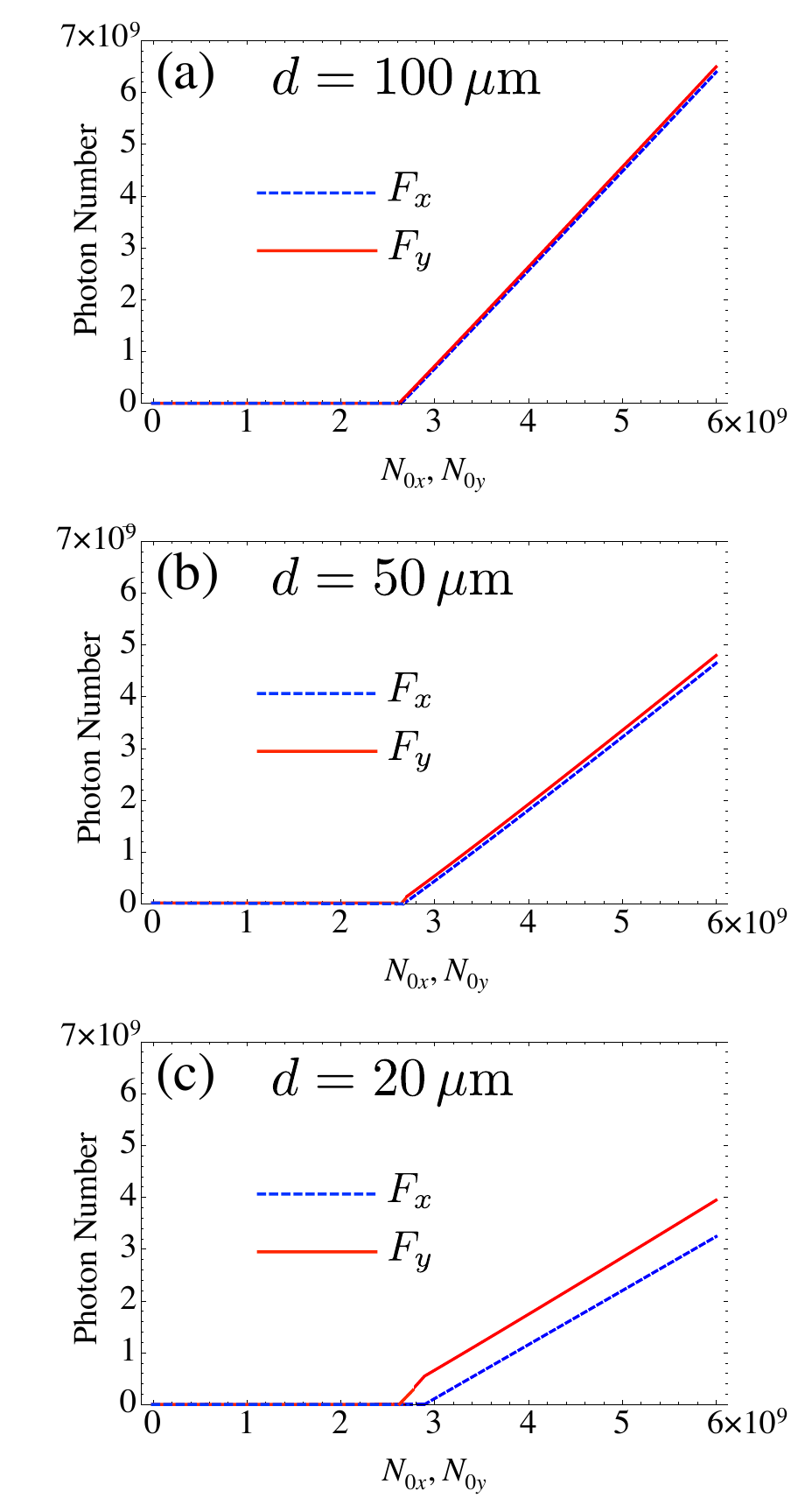}
\caption{(Color online) Evolution of the steady-state photon numbers $F_x$ and $F_y$ versus pumping rate $N_{0x}=N_{0y}$. The values of the parameters are $w=62\,\mu\mathrm{m}$, $1/\gamma_x=6\,\mathrm{ns}$, $1/\gamma_y=6.05\,\mathrm{ns}$, $1/\Gamma=3\,\mathrm{ns}$, $\varepsilon=0.02$, $\kappa=6.3\times 10^{-2}\,\mathrm{s}^{-1}$. (a) $d=100\,\mu\mathrm{m}$;  (b) $d=50\,\mu\mathrm{m}$; (c) $d=20\,\mu\mathrm{m}$. }\label{Fig06}
\end{figure}

The steady-state solutions for the laser intensities can be obtained by finding the steady-state solutions of Eqs.\;(\ref{Eq27N3}) and (\ref{Eq27N4}). In particular, we can plot the intracavity photon numbers for the two modes versus pumping rates $N_{0x}$ and $N_{0y}$, as shown in Fig.\ \ref{Fig06} with the same parameters as in Fig.\ \ref{Fig04}. In this case we took three different values of $d$, just like in Fig.\ \ref{Fig04}. The two modes have slightly different losses ($\gamma_{x}>\gamma_{y}$), but equal pumping rates ($N_{0x}=N_{0y}$). One can again see the influence of the mode competition on the resulting unbalance in photon numbers.

\section{Nonlinear coupling constant}
\begin{figure*}[]
\includegraphics[width=1.0\textwidth]{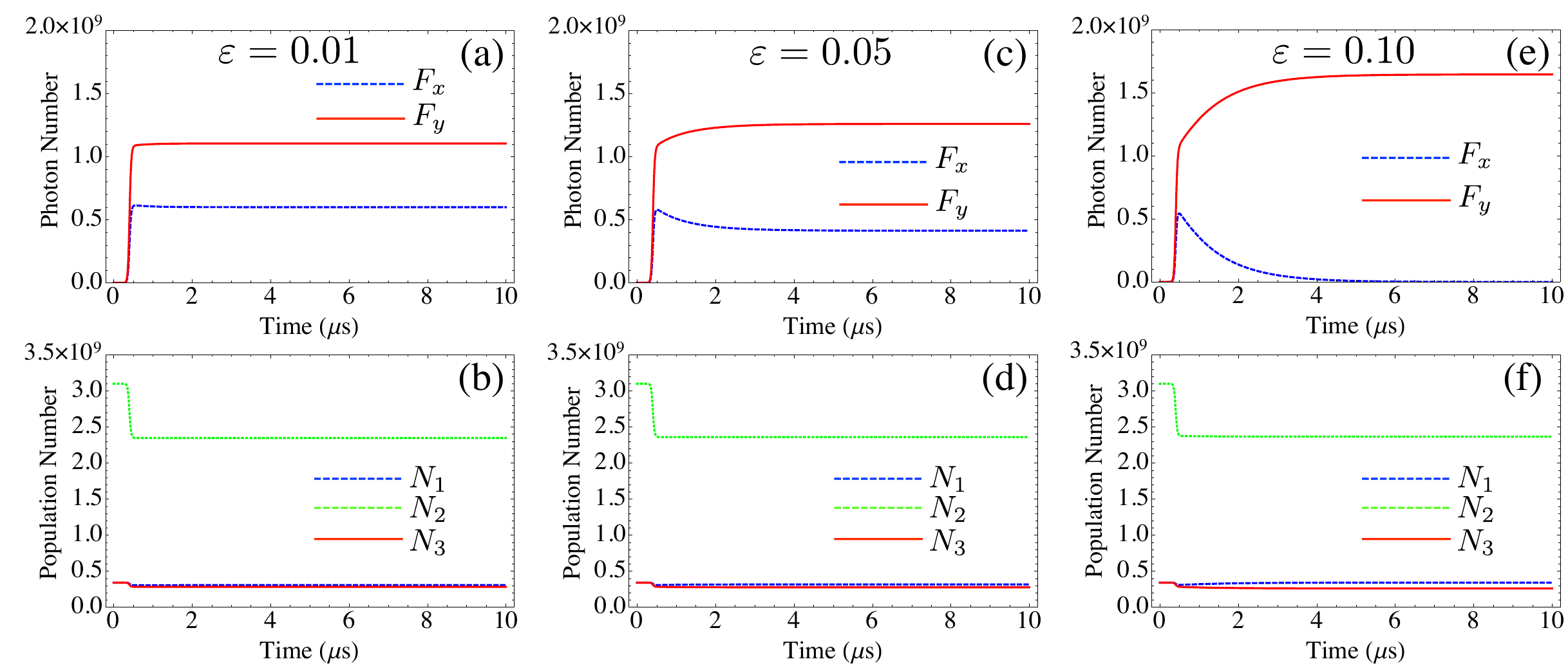}
\caption{(Color online) Time evolution of (a,c,e) the intra-cavity photon numbers $ F_{x} $ and $ F_{y} $ and (b,d,f) the population inversions $N_1$, $N_2$, and $N_3$ for $d=20\,\mu\mathrm{m}$, $w=62\,\mu\mathrm{m}$, $1/\gamma_x=6\,\mathrm{ns}$, $1/\gamma_y=6.05\,\mathrm{ns}$, $1/\Gamma=3\,\mathrm{ns}$, $\kappa=6.3\times 10^{-2}\,\mathrm{s}^{-1}$, $N_{0x}=N_{0y}=1.3\times\gamma_x/\kappa$. (a,b) $\varepsilon=0.01$; (c,d) $\varepsilon=0.05$; (e,f) $\varepsilon=0.1$.}\label{Fig07}
\end{figure*}
Let us consider the situation in which the laser obeys the class-A dynamical regime, as in Figs.\ \ref{Fig04} and \ref{Fig06}, which corresponds to the experimental results we wish to reproduce. The key phenomenon when we consider the simultaneous oscillation of the two modes is the nonlinear coupling, which has been experimentally investigated \cite{Pal2010}. In the case of weak saturation, Eqs.\ (\ref{Eq32}) and (\ref{Eq33}) show that the coupling does not only depend on the overlap $\eta$, as expected and as shown by the simulations of Fig.\ \ref{Fig04}. It also depends on the ratio $\varepsilon$ between the spin and the population relaxation rates. Fig.\ \ref{Fig07} shows the results obtained when one simulates the laser switch-on for different values of $\varepsilon$ when the two modes experience slightly different losses. This simulation is performed using Eqs.\ (\ref{Eq24}-\ref{Eq27}). It is clear from this figure that the larger $\varepsilon$, the stronger the coupling between the two modes. By increasing $\varepsilon$ from 0.01 to 0.1, one can see that a small difference between the losses $\gamma_x$ and $\gamma_y$ experienced by the two modes (which have the same gain) can lead either to a small difference in intensities [see Figs.\ \ref{Fig07}(a,b)] when $\varepsilon$ is small or to the fact that the stronger mode forbids oscillation of the other one [see Figs.\ \ref{Fig07}(e,f)] when $\varepsilon$ is larger.

\begin{figure}[]
\includegraphics[width=0.4\textwidth]{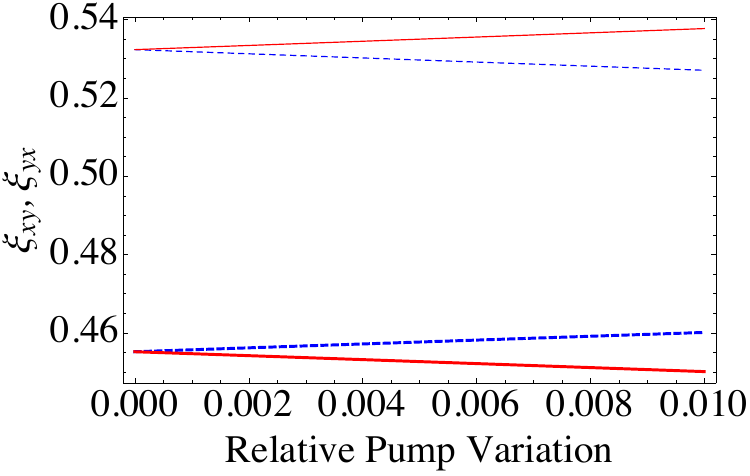}
\caption{(Color online) Evolution of $\xi_{xy}$ (dashed blue line) and  $\xi_{yx}$ (full red line) versus $\delta N_0$ with $N_{0x}=N_0(1+ \delta N_0)$ and $N_{0y}=N_0(1- \delta N_0)$. Thick lines: calculation from Eqs.\ (\ref{Eq38}) and (\ref{Eq39}) using the steady-state photon numbers $F_x$ and $F_y$ obtained from Eqs.\ (\ref{Eq27N3}) and (\ref{Eq27N4}). Thin lines: calculation from Eqs.\ (\ref{Eq32}) and (\ref{Eq33}). The values of the parameters are $d= 50\,\mu\mathrm{m}$, $w=62\,\mu\mathrm{m}$, $1/\gamma_x=6\,\mathrm{ns}$, $1/\gamma_y=6\,\mathrm{ns}$, $1/\Gamma=3\,\mathrm{ns}$, $\varepsilon=0.02$, $\kappa=6.3\times 10^{-2}\,\mathrm{s}^{-1}$, $N_0=1.3\gamma_x/\kappa$. }\label{Fig08}
\end{figure}

Eqs.\ (\ref{Eq32}) and (\ref{Eq33}) also predict  that if $ N_{0x}\neq N_{0y} $, i.e., if the pumping or equivalently the gain to loss ratios for the two modes are not identical, then the two coupling coefficients $\xi_{xy}$ and $\xi_{yx}$ are not identical as soon as $ \varepsilon $ is different from 0. To check whether this prediction remains valid beyond the third order approximation, we generalize Eqs.\ (\ref{Eq32}) and (\ref{Eq33}) using the following definitions \cite{Pal2010}:
\begin{eqnarray}
\xi_{xy}=-\dfrac{\partial F_x/\partial \gamma_y}{\partial F_y/\partial \gamma_y}\ ,\label{Eq38}\\
\xi_{yx}=-\dfrac{\partial F_y/\partial \gamma_x}{\partial F_x/\partial \gamma_x}\ .\label{Eq39}
\end{eqnarray}
Fig.\ \ref{Fig08} represents the evolutions of $\xi_{xy}$ and  $\xi_{yx}$ versus $\delta N_0$ when one varies the pumping rates of the two modes in two opposite ways, namely:
\begin{eqnarray}
N_{0x}=N_0(1+ \delta N_0)\ ,\label{Eq40}\\
N_{0y}=N_0(1- \delta N_0)\ .\label{Eq41}
\end{eqnarray}

Fig. \ref{Fig08} compares the results obtained from Eqs.\ (\ref{Eq38}) and (\ref{Eq39}) using the steady-state photon numbers $F_x$ and $F_y$ obtained from Eqs.\ (\ref{Eq27N3}) and (\ref{Eq27N4}) with the result obtained using the weak saturation approximation [Eqs.\ (\ref{Eq32}) and (\ref{Eq33})]. One can see that both calculations indicate that the coefficients $\xi_{xy}$ and  $\xi_{yx}$ become asymmetric when the pumping rates for the two modes become different ($N_{0x}\neq N_{0y}$). However, although the values of $\xi_{xy}$ and  $\xi_{yx}$ given by the weak saturation approximation [Eqs.\ (\ref{Eq32}) and (\ref{Eq33})] are close to the ones obtained from the complete model [Eqs.\ (\ref{Eq27N3}) and (\ref{Eq27N4})], their evolutions with the asymmetry of the pumping rates are opposite. Besides, the difference betweeen $\xi_{xy}$ and  $\xi_{yx}$ leaves $C=\xi_{xy}\xi_{yx}$ unchanged, at least for small values of the pumping asymmetry $\delta N_0$. This can also be seen in the weak saturation limit using Eqs.\ (\ref{Eq32}) and (\ref{Eq33}):
\begin{equation}
 C =\xi_{xy}\xi_{yx}\simeq \frac{\eta^2}{4} (1 +2\varepsilon)\dfrac{(N_{0x}+N_{0y})^2}{N_{0x}N_{0y}}\ ,\label{Eq42}
\end{equation} 
which,  at first order, does not depend on $\delta N_0$ when $N_{0x}$ and $N_{0y}$ obey Eqs.\ (\ref{Eq40}) and (\ref{Eq41}).

Therefore, the observation of unequal values of the two coefficients $\xi_{xy}$ and  $\xi_{yx}$ that was found by  previous experiments \cite{Pal2010} can be explained in our model. Moreover, the model explains why these variations leave $C$ constant, as experimentally observed. 

\begin{figure}[h]
\includegraphics[width=0.4\textwidth]{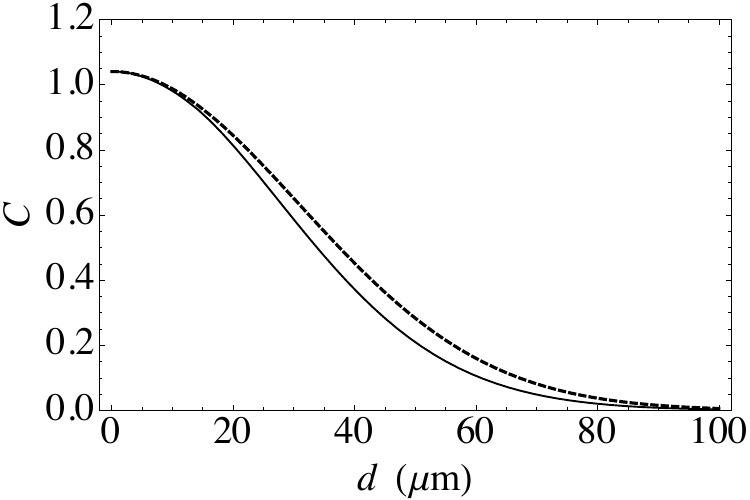}
\caption{ Evolution of $C=\xi_{xy}\xi_{yx}$ (full line) versus $d$ obtained from Eqs.\ (\ref{Eq38}) and (\ref{Eq39}) using the steady-state photon numbers $F_x$ and $F_y$ obtained from Eqs.\ (\ref{Eq27N3}) and (\ref{Eq27N4}). Dashed line: calculation from Eq.\ (\ref{Eq42}). The values of the parameters are $w=62\,\mu\mathrm{m}$, $1/\gamma_x=6\,\mathrm{ns}$, $1/\gamma_y=6\,\mathrm{ns}$, $1/\Gamma=3\,\mathrm{ns}$, $\varepsilon=0.02$, $\kappa=6.3\times 10^{-2}\,\mathrm{s}^{-1}$, $N_{0x}=N_{0y}=1.3\gamma_x/\kappa$. }\label{Fig09}
\end{figure}
Fig.\ \ref{Fig09} reproduces the evolution of $C$ with $d$. The results obtained either from the full calculation based on Eqs.\ (\ref{Eq38}) and (\ref{Eq39}) using the steady-state photon numbers $F_x$ and $F_y$ obtained from Eqs.\ (\ref{Eq27N3}) and (\ref{Eq27N4}), or from the simplified expressions of Eq.\ (\ref{Eq42}) lead to almost the same result, explaining the success of the simplified heuristic model \cite{De2013}. The maximum value of $C$, obtained for $d=0$, depends on $\varepsilon$. The fact that it is larger than unity explains why a partial spatial separation is necessary to obtain simultaneous oscillation of the two polarizations. Moreover, this predicts the existence of polarization bistability for $d$ close to 0. These calculations are in good agreement with experimental results \cite{Baili2009,Pal2010}.

\section{Intensity Noise Properties}
\begin{figure*}[]
\includegraphics[width=1.0\textwidth]{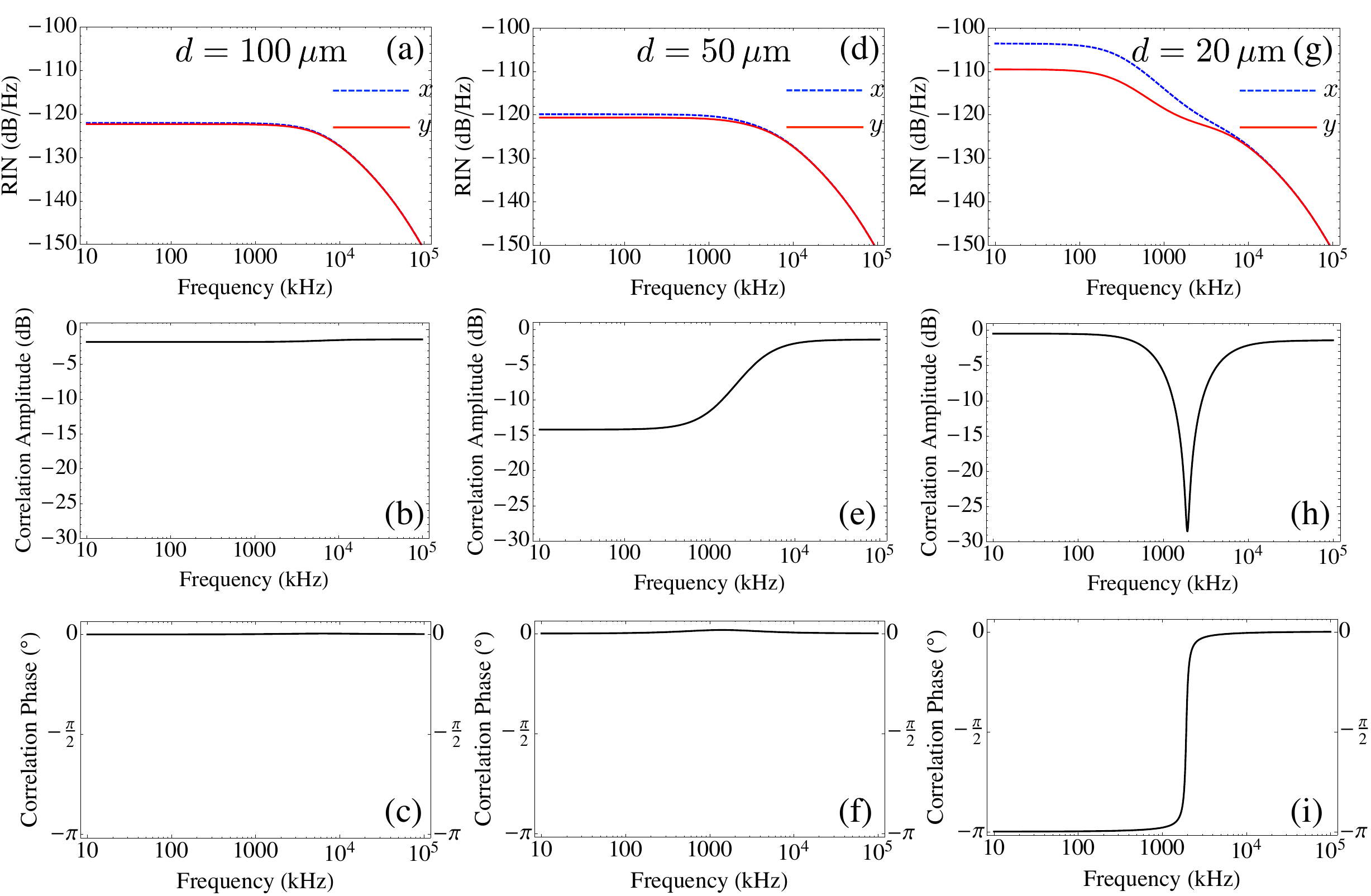}
\caption{(Color online) (a) Relative intensity noise, (b) correlation amplitude, and (c) correlation phase spectra calculated from the present model. The values of the parameters are $w=62\,\mu\mathrm{m}$, $1/\gamma_x=6\,\mathrm{ns}$, $1/\gamma_y=6.05\,\mathrm{ns}$, $1/\Gamma=3\,\mathrm{ns}$, $\varepsilon=0.02$, $\kappa=6.3\times 10^{-2}\,\mathrm{s}^{-1}$, $\overline{N}_{0x}=\overline{N}_{0y}=1.3\times\gamma_x/\kappa$, $\mathrm{RIN}_p=-135\ \mathrm{dB/Hz}$, and $d=100\,\mu\mathrm{m}$, corresponding to an overlap $\eta=0.074$. (d,e,f): Same as (a-c) for $d=50\,\mu\mathrm{m}$, corresponding to $\eta=0.52$. (g,h,i): Same as (a-c) for $d=20\,\mu\mathrm{m}$, corresponding to $\eta=0.90$.}\label{Fig10}
\end{figure*}
Since dual-frequency VECSELs are developed for microwave photonics applications \cite{Baili2009}, their noise properties are of particular interest. The simple heuristic model that has been used till now \cite{De2013} has led to a very good agreement with the measurements of intensity noise spectra and of intensity noise correlations between the two modes. Preceding work \cite{De2013} has also shown that, in the frequency range from 10 kHz to 100 MHz, the predominant source of intensity noise for the dual-frequency VECSEL originates from the pump intensity noise. This is an important issue for microwave photonics applications because this intensity noise is then transferred to the phase noise through the phase intensity coupling mechanism \cite{De2014}. The aim of the present section is thus to apply the present model, and more precisely Eqs.\ (\ref{Eq24}) to (\ref{Eq27N0}), to the description of the intensity noises of the two modes of the laser, including their correlations.

Since between 10 kHz and 100 MHz the dominant source of noise originates from the intensity fluctuations of the pump laser \cite{De2013}, we model this pump noise as follows:
\begin{eqnarray}
N_{0x}(t)=\overline{N}_{0x}+\delta N_{0x}(t)\ ,\label{Eq43}\\
N_{0y}(t)=\overline{N}_{0y}+\delta N_{0y}(t)\ ,\label{Eq44}
\end{eqnarray}
where $ \overline{N}_{0x} $, $ \overline{N}_{0y} $ are the average values of the unsaturated population inversions, which are proportional to the average pumping rates for the two modes, and $\delta N_{0x}(t)$ and $\delta N_{0y}(t)$ are their fluctuations. The fluctuations of the photon numbers of the two modes are defined as 
\begin{eqnarray}
F_{x}(t)=F_{x0}+\delta F_{x}(t)\ ,\label{Eq45}\\
F_{y}(t)=F_{y0}+\delta F_{y}(t)\ ,\label{Eq46}
\end{eqnarray}
where $F_{x0}$ and $F_{y0}$ are the steady-state solutions for the laser photon numbers and $\delta F_{x}(t)$ and $\delta F_{y}(t)$ are their fluctuations.
The pump fluctuations entering into the two partially spatially separated laser modes obey the following relations for frequencies $f$ between 10 kHz and 100 MHz:
\begin{eqnarray}
<\vert \widetilde{\delta N}_{0x}(f)\vert^2>&=&<\vert\widetilde{\delta N}_{0y}(f)\vert^2>\nonumber\\
&\equiv&<\vert\widetilde{\delta N}_{0}(f)\vert^2>\ ,\\
<\widetilde{\delta N}_{0x}(f)\widetilde{\delta N}_{0y}^*(f)>&=&<\widetilde{\delta N}_{0y}(f)\widetilde{\delta N}_{0x}^*(f)>\nonumber\\
&\equiv&\eta_p<\vert\widetilde{\delta N}_{0}(f)\vert^2>e^{i\psi_p},\end{eqnarray}
where the tilde holds for the the Fourier transformed variable, and $<.>$ holds for averaging. In our experiment, the pump relative intensity noise (RIN) level has been measured to be flat and equal to $-135\,\mathrm{dB/Hz}$ between 10~kHz and 100~MHz. The pump noise correlation factor $ \eta_{p}$ has been measured to be constant and equal to 0.85 over the same frequency range and the relative phase of the pump fluctuations $ \psi_p $ has been found to be equal to 0 \cite{De2013}. Therefore, the pump fluctuations entering into the laser modes are identical white noises, which are partially correlated ($ \eta_p<1 $) but are in phase ($ \psi_p=0 $).

By linearizing Eqs.\ (\ref{Eq24}-\ref{Eq27N0}) around their steady-state solutions with Mathematica, we are able to extract linear relations between the fluctuations of the photon numbers, of the population inversions, and of the pumping rate. After taking the Fourier transforms of these equations, we can eliminate the fluctuations of the population inversions and end up with a numerical expression relating the fluctuations of the photon numbers $\widetilde{\delta F}_{x}(f)$ and $\widetilde{\delta F}_{y}(f)$ in the frequency domain to $\widetilde{\delta N}_{0x}(f)$ and $\widetilde{\delta N}_{0y}(f)$. From these quantities we can obtain the relative intensity noise spectra (RINs) of the two laser modes:
\begin{eqnarray}
\mathrm{RIN}_{x}(f)=\dfrac{<\vert \widetilde{\delta F}_{x}(f)\vert ^2>}{F^2_{x0}}\ ,\\
\mathrm{RIN}_{y}(f)=\dfrac{<\vert \widetilde{\delta F}_{y}(f)\vert ^2>}{F^2_{y0}}\ .
\end{eqnarray}

The relative intensity noise of the pump is related to $\widetilde{\delta N}_{0}(f)$ by:
\begin{equation}
\mathrm{RIN}_{p}(f)=\dfrac{<\vert \widetilde{\delta N}_{0}(f)\vert ^2>}{N^2_{0}}\ ,
\end{equation}
where we have assumed that $ \overline{N}_{0x}= \overline{N}_{0y}\equiv N_{0} $, i.e., identical pumping for the two modes. 

The normalized spectrum of the correlation between the intensity noises of the two modes is defined as 
\begin{equation}
\Theta(f)=\dfrac{<\widetilde{\delta F}_{x}(f)\widetilde{\delta F}^*_{y}(f)>}{\sqrt{<\vert \widetilde{\delta F}_{x}(f)\vert^2> <\vert \widetilde{\delta F}_y(f) \vert^2>}}\ .
\end{equation}
$ \Theta(f) $ is a complex function having an amplitude and a phase. The correlation amplitude and phase spectra are respectively defined by $ \vert \Theta(f) \vert^2 $ and $ \arg[\Theta(f)] $. One has $ \vert\Theta(f)\vert\leq 1 $, where equality describes perfect correlation. 

Fig. \ref{Fig10} reproduces the results obtained for three values of the spatial separation $d$, namely 100, 50, and 20 $\mu$m. These results are in excellent agreement with the corresponding experiments and with the spectra obtained from the simple heuristic model that we had previously developed \cite{De2013}. In particular, these results reproduce the fact that for the weak coupling ($ C \simeq 0 $, see Fig.\ \ref{Fig09}) corresponding to $d=100\,\mu\mathrm{m}$ [see Figs.\ \ref{Fig10}(a-c)], the relative intensity noise (RIN) spectra look like the transfer function of first order filters and hence illustrate the class-A dynamical behavior of our laser. The RIN spectra for the two modes are not identical as we take slightly different losses for the two modes. The intensity noises for the two modes are partially correlated [see Fig.\ \ref{Fig10}(b)] and their correlated parts are in phase [see Fig.\ \ref{Fig10}(c)] all over the considered frequency range (10 kHz - 100 MHz). 

The simulation results for spatial separation $ d =50 \,\mu$m between the modes are shown in Figs.\ \ref{Fig10}(d-f). The two modes are moderately coupled ($ C \simeq 0.3 $) in this situation as shown by Fig.\ \ref{Fig09}. The RIN spectra again look like the transfer function of first order filters, thus confirming again the class-A dynamical behavior of the laser [see Fig.\ \ref{Fig10}(d)]. As shown by Figs.\ \ref{Fig10}(e,f), the correlation amplitude is very low for frequencies lower than $ \sim $1 MHz, whereas the correlation amplitude is strong and the correlation phase is 0 for higher frequencies, as observed experimentally in this situation \cite{De2013}.

Finally, Figs.\ \ref{Fig10}(g-i) reproduce the simulation results for a spatial separation $ d= 20\, \mu$m between the modes, which corresponds to a quite strong  nonlinear coupling ($ C \simeq 0.8 $) as reported in Fig.\ \ref{Fig09}. The RIN spectra as shown in Fig.\ \ref{Fig10}(g) now exhibit a change of slope at about 1~MHz . The correlation amplitude is strong except for a dip around 1~MHz, as shown in Fig.\ \ref{Fig10}(h). The correlation phase jumps from $ \pi $ at frequencies lower than 1 MHz to 0 for higher frequencies, as reported in Fig.\ \ref{Fig10}(i). 

This behavior, and more particularly the fact that the correlation phase exhibits only two values, namely 0 and $\pi$, has already been physically explained \cite{De2013}. It originates from the fact that the two linear orthogonal polarizations are coupled and hence there are two eigen-relaxation mechanisms for the out-of-equilibrium fluctuations of the whole system \cite{Otsuka1992}: an in-phase relaxation mechanism and an anti-phase relaxation mechanism. The in-phase response is independent of the coupling, whereas the anti-phase response strongly depends on the coupling and has a cut-off frequency ($ \sim $1 MHz for our VECSEL) much lower lower than the in-phase one. For $ d=100 \, \mu$m, the two laser modes are almost uncoupled ($ C\simeq0 $). As a result, the in-phase response dominates over anti-phase response for all frequencies within 10 kHz to 100 MHz and thus giving rise to partially correlated [see Fig.\ \ref{Fig10}(b)] and in phase [see Fig.\ \ref{Fig10}(c)] intensity noises induced by partially correlated ($ \eta_{p}<1 $), in phase ($ \psi_{p}=0 $) pump fluctuations. For $ C\simeq 0.3 $ in the case where $ d = 50\,\mu $m, the amplitude of anti-phase response becomes comparable to amplitude of the in-phase response for frequencies lower than the cut-off frequency of anti-phase response ($ \sim $1 MHz), whereas for higher frequencies the in-phase response dominates. This explains the low values of the correlation amplitude for frequencies lower than 1 MHz, and the high values of the correlation amplitude and 0 correlation phase for frequencies higher than 1 MHz [see Figs.\ \ref{Fig10}(e,f)]. For $ d=20\,\mu $m, which corresponds to  a stronger coupling situation ($ C=0.8 $), the anti-phase response strongly dominates over the in-phase response for frequencies lower than anti-phase cut-off frequency (1 MHz), but the in-phase response always dominates for higher frequencies. As a result, the correlation phase is $ \pi $ for frequencies lower than 1 MHz but $ 0 $ for higher frequencies [see Fig.\ \ref{Fig10}(i)]. Moreover, the phase jump at about 1 MHz [see Fig.\ \ref{Fig10}(i)] due to the transition of the laser dynamics from dominant anti-phase to in-phase behavior gives rise to the dip in the correlation amplitude [see Fig.\ \ref{Fig10}(h)] as the fluctuations of nearly identical amplitudes of the two modes interfere destructively. 

\section{Conclusion}
In this paper, we have developed a rate equation model taking the spin dynamics of the carriers into account, in order to predict the different properties of dual-frequency VECSELs sustaining oscillation of two linear orthogonal polarization modes. Specifically, we discussed how the relative value of the spin relaxation rate of the carriers with respect to the carrier decay rate inside the QW-based gain medium determines the dynamics of the dual-frequency VECSEL. This model has proved to be successful in describing the steady-state and transient intensities of the laser, the behavior of the nonlinear coupling between the two modes, and also the properties of the intensity noises and their correlations. This theoretical approach to the dual-frequency VECSEL, based on the ideas of SFM model, on the one hand generalizes the previous simple heuristic model reported in Refs. \cite{De2013, De2014}, and on the other hand predicts a few properties of dual-frequency VECSELs such as the possible existence of bistability of the two modes and unequal values of the ratios of cross- to self-saturation coefficients, which could not be explained by the simple model \cite{Pal2010}. Moreover, it opens interesting perspective concerning the dynamics of the laser when the beat frequency between the two modes becomes too slow to be averaged out, leading to possible self-pulsing and other dynamical phenomena \cite{Brunel1999a,Brunel1999b}.

\section*{Acknowledgments}
The authors are happy to thank the reviewer of this paper for his/her remarks that have strongly helped to improve this work. This research was partially supported by the Agence Nationale de la Recherche (Project NATIF No. ANR-09-NANO-012-01).

\end{document}